# Charge-Transfer Dynamics and Nonlocal Dielectric Permittivities Tuned with Metamaterial Structures as Solvent Analogues

Kwang Jin Lee[1], Yiming Xiao[1,2], Jae Heun Woo[1,3], Eunsun Kim[1], David Kreher[2], André-Jean Attias[2], Fabrice Mathevet[2], Jean-Charles Ribierre[1]*[†], Jeong Weon Wu[1]*, Pascal André[1,4]*

**Charge transfer (CT) is a fundamental and ubiquitous mechanism in biology, physics and chemistry. Here, we evidence that CT dynamics can be altered by multi-layered hyperbolic metamaterial (HMM) substrates. Taking triphenylene:perylene diimide dyad supramolecular self-assemblies as a model system, we reveal longer-lived CT states in presence of HMM structures with both charge separation and recombination characteristic times increased by factors of 2.4 and 1.7, i.e. relative variations of 140 and 73 %, respectively. To rationalise these experimental results in terms of driving force, we successfully introduce image dipole interactions in Marcus theory. The non-local effect herein demonstrated is directly linked to the number of metal-dielectric pairs, can be formalised in the dielectric permittivity, and is presented as a solid analogue to local solvent polarizability effects. This model and extra PH3T:PC60BM results show the generality of this non-local phenomenon and that a wide range of kinetic tailoring opportunities can arise from substrate engineering. This work paves the way toward the design of artificial substrates to control CT dynamics of interest for applications in optoelectronics and chemistry.**

Charge separation (CS) and recombination (CR) are two sides of the same coin, which impact numerous applied and fundamental phenomena,[1,2] in electronics,[3,4] magnetism,[5] biology and photosynthesis,[6] as well as chemistry.[7,8] Understanding and artificially tuning charge transfer (CT) dynamics are then of the greatest importance. Optical spectroscopy techniques have equipped researchers with powerful tools to probe charge exchanges between donors and acceptors with sub-picosecond time resolution over a wide spectral range. Interfaces,[4,8] solvents,[9-11] as well as atomic or molecular organisation and engineering[12-21] are among the triggers which scientists investigate and rely on to gain control over CT dynamics.

Nanostructured substrates have emerged as a promising class of artificial materials to tune light matter interactions. In this system, plasmonic structures are receiving tremendous attention and have recently been used to induce strong coupling to control chemical landscapes and charge mobility in cavities and above a patterned metal film, respectively.[22-24] Metamaterials are richer component structures with alternating metal, dielectric or semiconductor materials whose dielectric permittivity are carefully combined.[25-29] They have proved to be relevant to numerous properties and applications including super-resolution imaging,[30,31] optical spin Hall effect,[32,33] non-linear optics,[34] holography,[35] super-absorption and light trapping,[36,37] as well as optoelectronic devices.[38] Hyperbolic metamaterials (HMMs), which display hyperbolic iso-frequency contour when dielectric permittivities have opposite signs in two principal optical axes, have attracted a lot of attention when combined with chromophores located near or inside these nanostructures with which usually unsupported high-$k$ modes can propagate up to a cut-off wavelength corresponding to the pair thicknesses.[29] HMMs have been shown to alter both angular dependent emission through electric and magnetic dipoles,[39,40] and radiative dynamics through increased photonic density of states.[41-44] It was also recently suggested that HMMs could affect photodegradation.[45] However, extending far beyond the current literature we demonstrate experimentally and theoretically that HMMs also bear unexplored potentials to trigger CT dynamics alterations in non-local dielectric environments. In this context, we also draw a distinction between HMM properties and structures, with the former being determined by the optical phase diagram[46] and the polarization[47] through the hyperbolic energy-momentum dispersion relation obtained in an effective medium description, whilst the latter herein investigated are independent of these parameters. This distinction is further commented in section SI-VI.2, and then in the present manuscript, the terms HMM structures and HMM substrates refer to the non-local effect resulting from multilayer stacks such as the one presented in Figure 1 |a1.

We focused on organic semiconductors because of their relevance to the growing field of organic electronics in which CT control is essential.[12-21] These materials are also well suited for molecular engineering, which provides opportunities to control the spatial organisation of donor:acceptor (D:A) substituents, as well as their relative highest occupied molecular orbital and lowest unoccupied molecular orbital (HOMO-LUMO) levels. For this, self-assembled systems of liquid crystalline dyad and multiad molecules with covalently linked D:A units are extremely attractive.[48-54] Under appropriate design and preparation conditions, they can indeed produce highly ordered self-assemblies with stacked π-molecules segregated in relatively D and A independent lamellar or columnar domains of controlled interfaces and nanometer sizes.[48-54] These characteristics make them ideal model systems in which charge separations and re-combinations are not influenced by domain size and interface distributions, which otherwise impact on where excitons form, how they diffuse, dissociate in CT states or free carriers and how recombination occurs in a geminate or non-geminate manner.

In this work, we took advantage of discotic D:A dyad supramolecular self-assemblies to investigate the influence of nanostructured substrates on CT dynamics. Both CS and CR characteristic times increased with the number of metal-dielectric pairs. We developed Marcus theory framework by inserting image dipole interactions in the dielectric permittivity to describe qualitatively the slower CT dynamics we observed near metamaterial interfaces. This approach paves the way to original artificial substrates designed to control CT dynamics, and will impact fields including photonics and chemistry, through applications such as optoelectronics, photovoltaics, and photocatalysis.

---

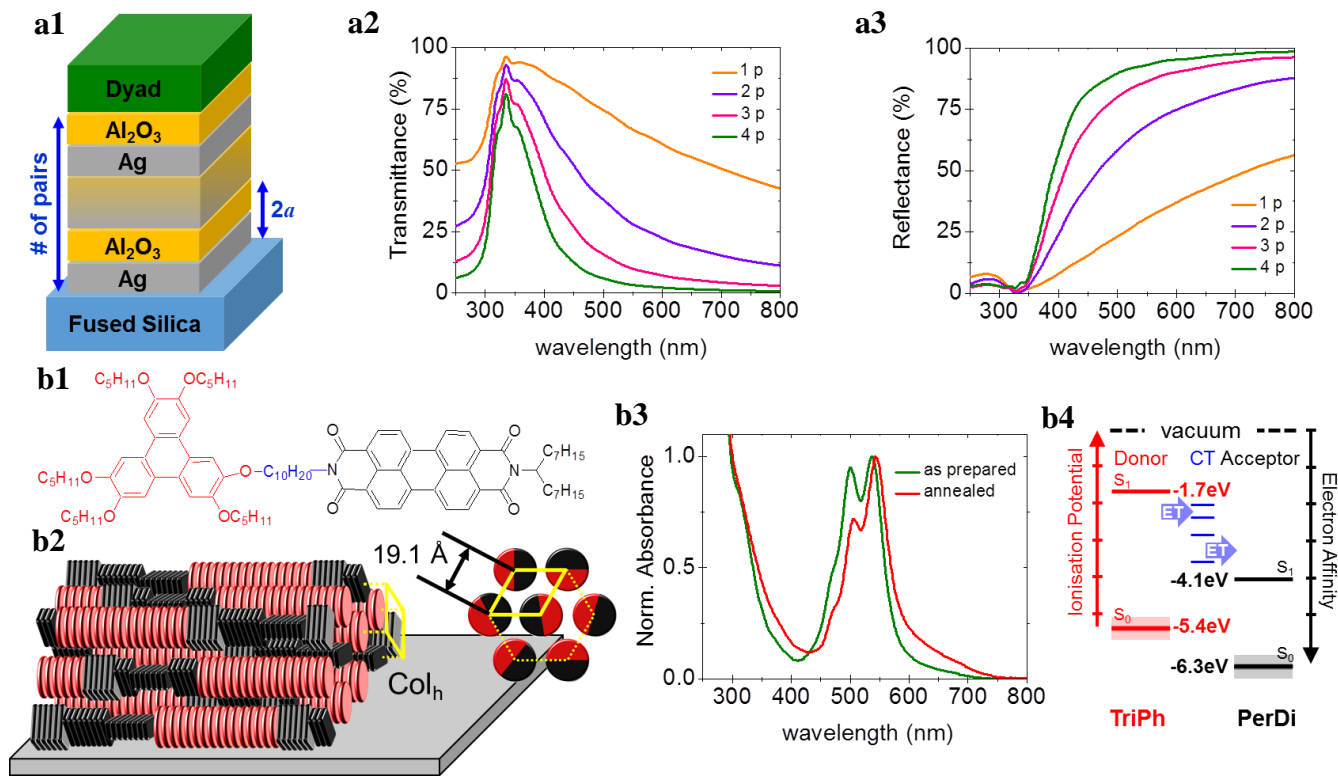

**Figure 1 | System under investigation. a**, Ag/Al$_2$O$_3$ multilayer HMM: Schematic representation (**a1**) including the number of pair (# p), the pair thickness and periodicity (2$a$ = 20 nm), the metal volume fraction ($f_v$ = 0.5). Calculated transmittance (**a2**) and reflectance (**a3**) as a function of wavelength for different number of pairs under normal incidence. **b**, Triphenylene-perylene diimide dyad molecules. **b1**, Chemical structure. **b2**, Schematic representation of the columnar structure planar orientation on Al$_2$O$_3$ surfaces evidenced by GIXS. **b3**, Steady state absorbance spectrum of as-prepared and annealed dyad thin films on fused silica substrates. **b4**, Schematic representation on the TriPh-PerDi energy diagram of charge transfer mechanisms in the dyad films.

### HMM and Organic semiconductor Properties

Figure 1 |a1 shows the multi-layered HMM structure used in this study with 10 nm thick silver and aluminium oxide alternative layers prepared as described in the methods. Figure 1 |a2 presents typical experimental HMM absorbance spectra with 1 to 4 number of pairs. The 325 nm dip is induced by the silver dielectric permittivity. As illustrated in Figure 1 |a2 and a3, increasing the metal amount with the number of pairs results in a decrease of the transmittance and an increase of the reflectance of the HMM. This leads to substrates that are almost perfect mirrors in the visible-NIR spectral range.

We focused on a triphenylene:perylene diimide dyad (TriPh:PerDi, Figure 1 |b1) composed of discotic mesogenic conjugated units providing suitable morphological and optoelectronic properties including stability under ambient conditions. Connected by a decyloxy flexible bridge, the TriPh (D) and PerDi (A) units are prone to self-assemble in a solid state multi-segregated D-A columnar structure at the local range.[53,54] Grazing Incidence X-ray Scattering (GIXS) of the dyad films on Al$_2$O$_3$ surfaces revealed a spontaneous planar orientation of columnar hexagonal structures (Col$_h$) consisting of domains of lying columns as illustrated in Figure 1 |b2 and SI-Figure 1|. The annealing step in the fluid state of the mesophase only develops the long-range order enhancing the regularity of structure and the planar orientation. The molecules and the columns are oriented edge-on and parallel to the surface of the substrate, respectively, and the average distance between the columns is ~ 19.1 Å. The conjugated core of the TriPh and PerDi columns are insulated from the Al$_2$O$_3$ substrate by a ~ 8 Å thick molten alkyl chain continuum. Figure 1 |b3 shows the absorption spectra of dyad thin films before and after annealing. The characteristics of both TriPh and PerDi are visible at high and low energy, respectively, and are broadened by π-π interactions. Annealing decreases the relative height of the PerDi high-energy peaks. This decrease of (0,1) vibronic peak relative to the (0,0) transition, at 491 nm and 527 nm respectively, is consistent with a decrease of the PerDi cofacial stacking of H-aggregate type. The relative growth of the absorbance above 600 nm occurring simultaneously is attributed to improved π-staking within the core of the columns. This aggregation contributes to the photoluminescence quenching of both moieties. Figure 1 |b4 shows that the relative positions of the TerPh and PerDi energy levels makes the former a natural electron donor and the latter a suitable acceptor. The most favourable CT mechanism is illustrated in the same figure,[55] with the donor photoexcitation followed by CS from the donor's LUMO to the acceptor's excited states.[54]

### CT Dynamics & Transient Absorption (TA) Spectroscopy

TA is a well-established technique to gain insight into CT phenomena. Figure 2 |a presents the TA dynamic spectra of an annealed dyad film. Considering the absorbance spectra of the D and A units, pumping at 325 nm mostly results in the TriPh photoexcitation. The weak negative response ($\Delta T/T > 0$) observed at high energy is associated with the PerDi ground-state bleaching. In contrast, the strong positive feature ($\Delta T/T < 0$) observed for $\lambda > 650$ nm corresponds to the PerDi anion absorption.[54,56-58] Focussing on this 725 nm signal, Figure 2 |b and c display the time resolved relative transmission variation for short (0-2 ps) and long (0-850 ps) delay times, respectively, obtained with annealed dyad thin films on a range of HMM substrates. The CT dynamics are summarised in Supplementary Table 2. The short delay time signal is associated with the formation of the CT state, i.e. the charge separation occurring when photogenerated excitons lead to Coulomb bound anionic PerDi and cationic TriPh. A $\Delta T/T$ plateau is reached within 2 ps (Figure 2 |b). In contrast, $\Delta T/T$ signal recovers, i.e. the CT state disappears over a longer time scale through geminate charge recombination, following an expo-

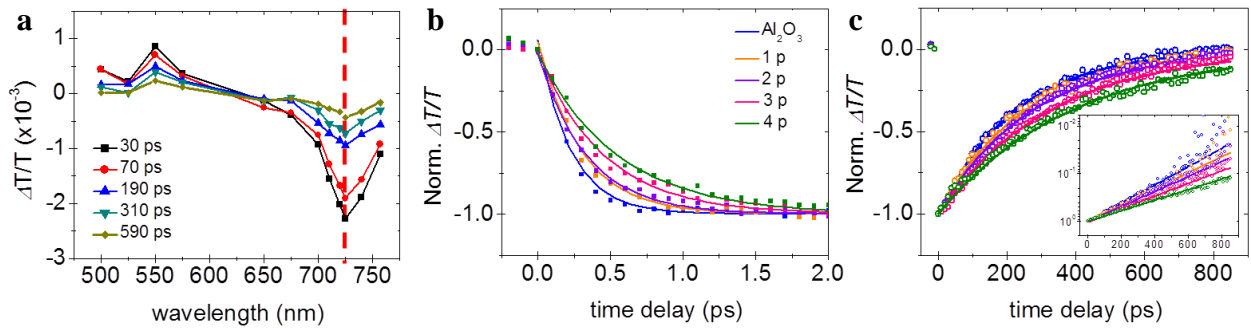

**Figure 2 | Transient absorption measurements of annealed dyad molecule thin films. a,** Relative transmission variation as a function of the wavelength for time delays ranging from 30 to 590 ps measured on fused silica with 325 nm pumping. Time resolved $\Delta T/T$ spectra probed at 725 nm monitoring **b**, the formation and **c**, the disappearance of the anionic PerDi (inset and SI: semilog plots with inverted scale illustrating the single exponential recovery behaviour). Colored lines are guide-for-the-eyes **(a)** and exponential fits **(b, c)**.

nential behaviour (Figure 2 |c-inset). Increasing the HMM number of pairs slows down the CT dynamics.

Noticeably, CS and CR dynamics of dyad films deposited on $SiO_2$ or $Al_2O_3$ interfaces are identical within the experimental precision (Figure 2 |a). In all cases, single exponential characteristic times described properly the dynamics (Figure 3 |a-b). On $Al_2O_3$ substrates, the CS kinetic is of the order of a few tenths of ps, which is consistent with the literature values of donor:PerDi systems.[57,58] CS is systematically faster by two-to-three orders of magnitude than CR. The dyad film CS is ~ 2.4 time (i.e. 140 %) slower on the 4-p HMM substrate than it is on-top of a plain $Al_2O_3$ interface, i.e. $p = 0$. These results were confirmed in transient reflectance mode, as presented in SI-Figure 6.

Similarly, CR time is increased by a factor 1.7 (i.e. 73 % slower). This is also far beyond the experimental uncertainty and clearly from the PerDi to the HMM substrates can be ruled out as it demonstrates the strong influence of the nanostructured substrates, which is consistently linked to the pair number. Electron transfer would have resulted in an accelerated CR. The alkyl chain melt is likely to prevent such CTs and the number of pairs would not have mattered. Furthermore, CT dynamics of the TriPh-PerDi dyad was shown to be independent of the fluence, which in the present study was explored up 45 μJ/cm². Presented in section SI-III.3.c, this fluence investigation reveals that the dyad molecules provide a good model system to probe the CT states, with all the measurements herein presented being below the threshold above which exciton-exciton annihilation, exciton-charge annihilation, and bimolecular charge recombination occur. Consequently, excitation density effects can be ruled out and a more exotic interpretation needs to be considered. The CS to CR ratios are displayed in Figure 3 |c. Whilst the resulting experimental uncertainties are increased, we note that the CS to CR time ratio appears tlingo also be a function of the pair number, which suggests that non-trivial phenomena control the HMM alteration of the CT dynamics. The CS and CR slowdowns were also observed in equimolar blends of TriPh and PerDi moieties (SI-Figure 7|; CS:CR blend slowdown ~47:14 % compared with 150:62 % for the dyads). Consequently, self-assembly enhances the nanostructured substrates CT dynamic effects, which are nonetheless present in systems of lower D-A segregation levels. In contrast, this effect could not be observed only by increasing the thickness of a single metal layer as shown in SI-Figure 8. As a result, Figure 3 | clearly establishes that HMM substrates can be used to increase and then control the CS and CR characteristic time constants in organic semiconductors.

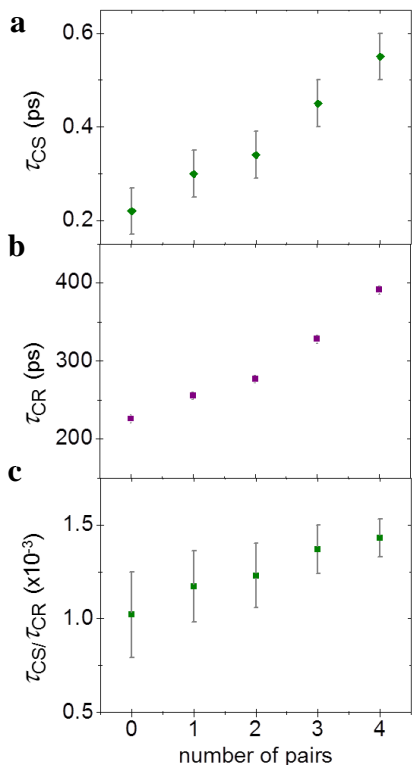

**Figure 3 | Quantitative TA analysis.** Characteristic times of charge **a**, separation and, **b**, recombination. **c**, Separation to recombination time ratio in dyad annealed thin films as a function of the HMM number of $Ag/Al_2O_3$ pairs.

**Marcus Theory Development with Image Dipole Interactions**

Marcus theory describes non-adiabatic CT reaction rates, $k_{CT}$, in terms of activation Gibbs free energy, $\Delta G^*_{CT}$, electronic coupling between the initial and final states (CT integral), $V_{DA}$, total reorganisation energy, $\lambda$, and thermal energy, $kT$:[9-11]

$$k_{CT} = (\frac{4\pi^3}{h^2 \lambda kT})^{1/2} |V_{DA}|^2 \exp(-\frac{\Delta G^*_{CT}}{kT}) \quad (1)$$

$\Delta G^*_{CT}$ is linked to $\lambda$ and to the Gibbs free energy gain, $\Delta G_{CT}$ also called driving force, through a parabolic relation.

$$\Delta G^*_{CT} = \frac{(\lambda + \Delta G_{CT})^2}{4\lambda} \quad (2)$$

It predicts a maximum of the CT rates for $-\Delta G_{CT}=\lambda$ (barrierless point). In the 'normal' region ($-\Delta G_{CT}<\lambda$) any driving force increase (more negative $\Delta G_{CT}$) lowers the activation energy barrier, hence, increases the reaction rate. In the 'inverted' ($-\Delta G_{CT}>\lambda$) region, the activation energy reappears and a driving force increase reduces the CT rate. As detailed in the SI, we introduced in an extended Marcus theory framework CT state dipoles and their image potentials in the HMMs. From this general formalism, we isolated a perturbation energy term,

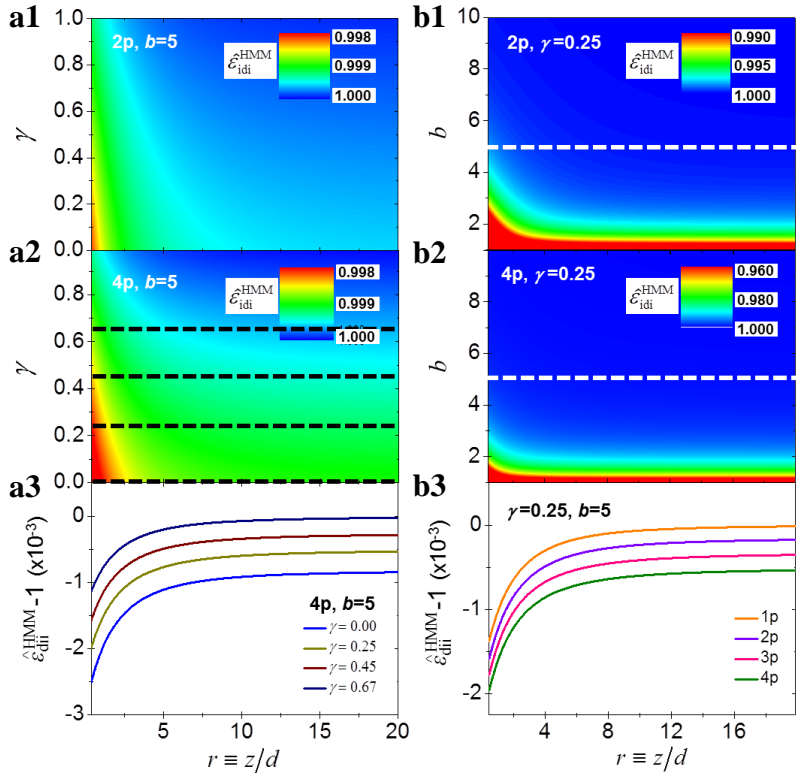

**Figure 4 | Calculated non-local dielectric permittivity manipulated by HMM image dipole interactions.** False-color plots as a function of the normalized dipole-HMM interface distance ($r$) and **a**, the orientation ($\gamma$) of the dipole, **b**, the normalized HMM structure parameter ($b$) for two (**1**) and four pairs (**2**). The dashed line corresponds to cuts at selected $\gamma$ and $b$ values presented in (**3**). $z$ : distance from the top $Al_2O_3$ HHM interface, $d$ : length of the dipole. $\gamma = (1, 0, 2/3)$ : dipoles parallel, perpendicular and isotropically oriented compared to the HHM surface. $b = a/d$ : normalized dipole length-HMM thickness layer, in the present dyad system $b \approx 5$.

$$\delta G_{CT}^{HMM} = \Delta G_{CT}^{HMM} - \Delta G_{CT}^{Al_2O_3} \quad (3)$$

The logarithm of the HMM to $Al_2O_3$ CT reaction rate ratio contains a constant and both $\delta G_{CT}^{HMM}$ linear and quadratic terms.

$$\ln \frac{k_{CT}^{HMM}}{k_{CT}^{Al_2O_3}} \cong -\alpha_{CT}\left[2(\lambda_{CT} + \Delta G_{CT}^{Al_2O_3})\delta G_{CT}^{HMM} + \delta G_{CT}^{HMM\,2}\right] \quad (4)$$

with $\alpha_{CT} = (4\lambda_{CT}kT)^{-1}$. Equation (4) applies to both CS and CR. The perturbation induced by the HMM can be written as

$$\delta G_{CS-\gamma}^{HMM} = -\delta G_{CR-\gamma}^{HMM} = -\Delta\Phi_\gamma^{HMM} \frac{e^2}{4\pi\varepsilon_0}\Delta_{1/R} \quad (5)$$

where $\gamma$ is the dipole orientation parameter, $\Delta_{1/R}$ has the dimension of the inverse of a distance (SI-IV.1) and is not expected to be drastically altered by the substrates. $\Delta\Phi_\gamma^{HMM}$ value depends on the HMM image dipole interactions manipulated dielectric permittivity, $\varepsilon_{idi}$, linked to CT Coulomb interactions (SI-IV.2): any CT effect resulting from HMM substrates is a function of $\varepsilon_{idi}$. Figure 4 |a-b show for 2 and 4 pairs that, at any dipole-HMM interface distance, $\hat\varepsilon_{idi}$ varies monotonously with dipole orientations and HMM structures. The stronger $\hat\varepsilon_{idi}$ variation corresponds to dipoles perpendicular to the HMM interface. This is further illustrated in Figure 4|a3. Increasing the thickness of the dielectric and metal bilayer (equivalently decreasing the dipole characteristic length) decreases $\hat\varepsilon_{idi}$ (Figure 4 |b1-2). The strongest effects should be observed for thin layers and large dipoles. The experimental data were obtained for $b=5$ for which Figure 4 |b3 shows the variation of $\hat\varepsilon_{idi} -1$ as a function of the normalised distance from the HMM interface: The larger the pair number, the larger the image dipole potential effect.

**Theoretical Analysis of the Experimental Data**

In Figure 5 |a, $\Delta\Phi_\gamma^{HMM}$ plateaus for $p>2$ when the dipole distribution is isotropic ($\gamma \sim 2/3$) but keeps increasing with the number of pairs when dipoles are more perpendicular to the substrate. This does not exclude a distribution of dipole orientation, however, perpendicular orientation will stabilise more dipoles than if $\gamma \neq 0$, consequently for the sake of simplicity the rest of the manuscript only considers dipoles with $\gamma = 0$.

Considering equation (4) naturally leads to neglect the 2$^{nd}$ order perturbation. Then equations (4) and (5) lead to a linear variation of the logarithm of the experimental CS ratios as a function of $\Delta\Phi$.

$$\ln \frac{k_{CS}^{HMM}}{k_{CS}^{Al_2O_3}} = -\ln\frac{\tau_{CS}^{HMM}}{\tau_{CS}^{Al_2O_3}} \cong \zeta_{CS}\Delta\Phi_{CS}^{HMM} + \text{const.} \quad (6)$$

with $\zeta_{CS} \equiv \frac{e^2}{4\pi\varepsilon_0}\frac{(\lambda_{CS} + \Delta G_{CS}^{Al_2O_3})}{2\lambda_{CS}kT}\Delta_{1/R}$.

Figure 5 |a-b were combined to obtain Figure 5 |c which validates the prediction set by equation (6). Whilst a small contribution of $\Delta\Phi^2$ cannot be excluded, it does appear to be smaller than $\Delta\Phi$'s at least for the experimental data accessible with this system. Noticeably, the slope on Figure 5 |c is positive, meaning that $\zeta_{CS}$ should be negative. As the term $\lambda+\Delta G$ is positive in the normal region, $\Delta_{1/R}$ should be negative. Whilst it difficult to assess exactly the value of $\Delta_{1/R}$ its sign, obtained from a solution approximation, is indeed negative (SI-IV.1.a). This emphasizes the consistency between the model and the experimental data.

Experimentally, both CS and CR rates decrease with the pair number (Figure 3 |a-b). However, equation (5) shows that the HMM affects the CS and CR driving forces with the same amplitude and opposite signs. Consequently, CS is in the normal region, whereas CR is in the inverted region (Figure 5 |e1). It is then important to note that through the variation of the CT rates with the HMM pair number it becomes possible to identify the regions of the Marcus diagram in which a charge transfer occurs. The characteristic times of CS are orders of magnitude shorter than CR's, with a stronger variation of CS than CR (2.5 vs 1.7) when the dyad is on $p=4$ HMM substrates. The CS rate variation with HMM pairs is steeper than for CR, which should then stand in the inverted region and close to the barrierless point. There, $\lambda_{CR}+\Delta G_{CR}\approx 0$, so that equation (4) applied to CR should be:

$$\ln \frac{k_{CR}^{HMM}}{k_{CR}^{Al_2O_3}} \cong \zeta_{CR}.\Delta\Phi^{HMM\,2} + \text{const} \quad (7)$$

with $\zeta_{CR} \equiv \frac{1}{4\lambda_{CT}kT}(\frac{e^2}{4\pi\varepsilon_0}\Delta_{1/R})^2$.

Figure 5 |d shows the resulting curves for $\gamma=0$. The $\Delta\Phi$ quadratic dependence is verified with a positive slope as expected from the $\zeta_{CT}$ sign. Note that the CR data were also plotted as a function of $\Delta\Phi$ in SI-Figure 14|, the mismatch with a linear variation reinforces the consistency of the arguments developed above. It is then worth noticing that whether the variation of the CR rates is linear or quadratic with $\Delta\Phi^{HMM}$ can be used to identify whether or not a charge transfer occurs near the barrierless point.

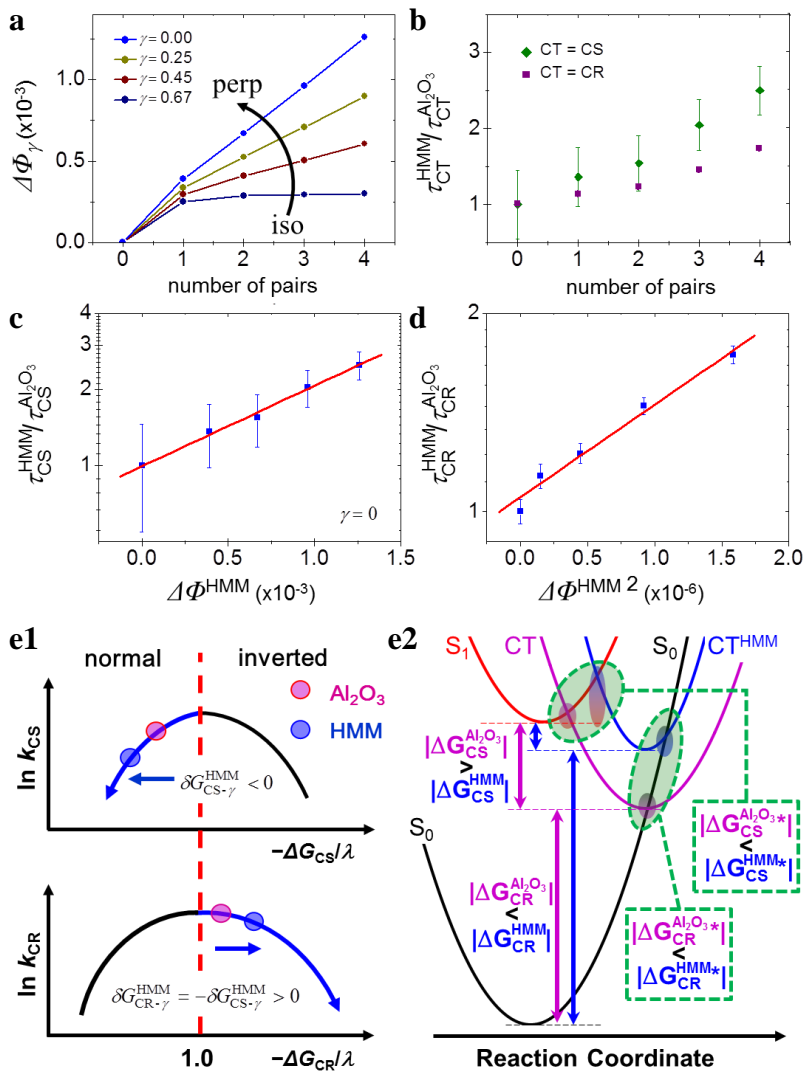

**Figure 5 | Modelling and Schematics. a**, Variation of the integrated dimensionless parameter, $\Delta\Phi$, from 1 to 4 pairs and for selected dipole orientations. **b**, Experimental characteristic time ratios of the CT state formation (◆) and disappearance (■) as a function of the number of pairs. **c**, CS and **d**, CR characteristic time ratios as a function of $\Delta\Phi$ and $\Delta\Phi$ square, respectively, for $\gamma=0$. The straight lines are the fitted linear curves. **e1**, Dependence of the logarithm of the CT rates ($k_{CT}$) with the driving force ($-\Delta G$) to reorganisation energy ($\lambda$) ratio. The red dashed line represents the barrierless point. The purple and blue dots are associated with standard and HMM substrates, respectively. The arrow indicates the impact on $-\Delta G/\lambda$ and $k_{CT}$ of the HMM substrate structure. **e2**, Marcus theory energy parabolas as a function of the nuclear coordinates of the reagents and products; $S_0$ and $S_1$ are the ground and first excited states of the donor, while CT and $CT^{HMM}$ are the charge transfer states of the dyad on standard and HMM substrates, respectively ; the colored arrows and ellipses are associated with the CS and CR energy gains ($|\Delta G|$) and activation energies ($|\Delta G^*|$), respectively. Due to the Marcus parabola displacement under the effect of HMM substrates, the energy gain of CS is decreased and CR is increased (arrows and equations on the right-hand side), whilst both CS and CR activation energies are increased (ellipses and equations on the left hand-side).

Finally and to complete the Marcus theory description, Figure 5 |e2 suggests a possible projection on a simplified energy parabola diagram of the CT alteration mechanisms in presence of HMM substrates. The limitations of this simplified representation are discussed in SI-VI.3. CS corresponds to transfers from the red to the purple and blue curves. CS occurs in the normal region and its activation barrier increases when the dyad is on the nanostructured substrates. Consequently, HMMs could move the CT parabola upwards. This is consistent with CR being in the inverted region and increasing its activation barrier with HMM structures. CR corresponds to a transfer from the purple/blue parabola to the black one.

## Summary, Discussion and Outlook

We experimentally evidenced that organic semiconductor CT dynamics depends on substrate nanostructures. CS and CR characteristic times were both increased by factors up to 2.5 and 1.7, i.e. slower by ~ 140 and 73 %, respectively. Marcus theory could qualitatively describe this trend when including Coulomb interactions and image dipoles in the multi-layered system to reveal the role of HMM image dipole interactions in manipulating dielectric permittivity with non-local effects. The reported control of the CT dynamics is analogous to the use of solvents presenting different dielectric permittivities. This solvent effect is well known,[9-11] but, to the best of our knowledge, unheard of in solid HMM nanostructured substrates. Furthermore, the relatively small variation of the non-local dielectric permittivity (~4 %) compared to a more traditional solvent effect needed to get a comparable CT dynamic rate alteration is attributed to the very different mechanisms involved. In the former case, directional image dipole interactions apply to a D:A thin films, whereas the latter relies on variation of the polarizability of the solvent molecules surrounding each and every single of the D:A moieties.

With the present results, original and interesting opportunities arise in terms of substrate design to control CT dynamics. In TriPh:PerDi systems, CS was slowed down and occurred in the normal region. However, we anticipate that by preserving $\Delta_{1/R}<0$, but selecting a system such that CS occurs in the inverted region, CS and CR will accelerate and slow down, respectively, leading to an even longer CT overall lifetime. Creating such situations extends beyond the scope of this work and would need non-trivial molecular engineering such as adjusting exciton delocalisation and reorganisation energies. Nonetheless, we anticipate the consequence of the present findings to be that any system having CS in the normal region and CR in the inverted region will see both CT rates slowed down near an HMM substrate. However, if CR is in the normal region then its dynamics will be accelerated and similarly if CS is in the inverted region then its dynamics will also be accelerated instead of being slowed down. This is set by the direction of the arrows and the bell shape of the Marcus diagram in Figure 5e1|. This situation paves the way towards the design of molecular systems and the exploration of their response when located near an HMM substrate. Exploring the limit of the present model and data, we note that this D:A system is dominated by CT processes, which along with π-stacking contribute to quenching the fluorescence. It would be relevant to identify and study systems in which a fair competition between CT and photoluminescence could be observed. A crossover should occur as the former slows down when tuning the dielectric permittivity with image dipole interactions, while the latter accelerates when increasing the photonic density of states.

Marcus theory finds applications in numerous systems from solutions to interfaces and solids. The present work represents an original extension of it towards solid multi-layered media. Image dipole interactions are influenced by dyad-HMM interface separations, dipole orientations, charge separation within the dipole, as well as HMM periodicity and pair number. From this general features, one can then envisage that more and thinner

HMM layers should induce a stronger CT dynamic alteration. Similar behaviours are expected for any nanostructured media allowing the periodic formation of image dipoles. We focussed on layered $Ag:Al_2O_3$ but other inorganic and organic materials as well as dissymmetric substrates should provide some control over CT dynamics. The generality of this finding is demonstrated not only by the model we developed but also by the results obtained with P3HT:PCBM, which to remain concise are discussed with SI-Figure 16|. We used organic semiconductors to evidence and demonstrate the impact of HMMs on CT dynamics, however, such phenomena should also occur in inorganic semiconductor materials. Extensions of this work should consider rod based HMMs as well as the design of structures increasing or modulating $\Delta\Phi$ in space for instance to create CT dynamic gradients or induced chirality in chemical reactions.

CT reactions are indeed ubiquitous across systems including biology, photonics, optoelectronics as well as chemistry to name but a few.[1-8] In this context, the present results pave the way towards an original strategy to tailor chemical reaction dynamics beyond the primary CT we have focused on. We anticipate that this type of substrate bares high potentials for the design of original optoelectronic devices, in which the recombination of CT states or even free charges could be delayed, for instance to extract charges as the case in photovoltaic applications.

In conclusion, we successfully evidenced that multi-layered HMM substrates could control organic semi-conductor CT dynamics. Demonstrating this finding generality, this non-local effect of metamaterials was shown with D-A dyad molecules and P3HT:PCBM, the model system of organic solar cells. We rationalised the experimental results within an extended Marcus theory framework, in which image dipole interactions were accounted for. Using these interactions to manipulate the dielectric permittivity formalism allowed to tune CT driving forces near the HMM interface. Whilst the underlying mechanisms are fundamentally different, changing the number of metal-dielectric pairs was shown to provide a non-local trigger analogous to the local effect of immersing the molecules in solvents of different polarisabilities. HMM structures with different numbers of pairs made it possible to scan $\Delta G/\lambda$ in thin films and to identify the regions of the Marcus diagram in which CS and CR occur. This could be achieved without changing the chemical structure of the D-A material, and without affecting the compounds organisation and morphology. To the best of our knowledge, this is unheard of with other substrate structures and it opens further opportunities to gain insight into structural and electronic properties of D-A compounds. Both CS and CR time constants were shown to be influenced and the model we developed suggests that both molecular and substrate engineering could offer a much wider range of situations. Consequently, this work not only unveils so far hidden opportunities to design artificial substrates allowing further control of CT kinetics, but will also find applications in fields as diverse as photonics, optoelectronic devices and chemistry.

### Methods.
**Sample preparation**. The multi-layered substrates were deposited on fused silica by Korea Advanced Nano Fab Center (KANC). The deposition was completed by ion beam evaporation to form 10 nm thick metal and oxide successive layers with a volume fraction $f_v = 0.5$. Four pair HMM is the upper limit above which the transmitted probe beam used in the transient absorption experiments is too weak to be reliably measured, and which then sets the limit of the optical characterisation of the organic semiconductor thin films. The dyad molecules used in this work were synthesized and purified following a procedure already published and summarised in the SI. Cyclic voltammetry measurements of TriPh, PerDi and dyad were carried out in chloroform solution using 0.1 M tetrabutylammonium hexa-fluorophosphate. 40 nm thick films were formed by spin-coating (5 mg/mL in dichloromethane) on cleaned HMM substrates. The self-assembly of the dyad molecules was optimized by a short annealing treatment of 120 min at 120 °C completed on a hot-plate under ambient conditions. Regioregular poly(3-hexylthiophene) (P3HT, $M_w$ = 57 kDa, pdi = 2.9), the soluble fullerene [6,6]-phenyl-$C_{60}$ butyric acid methyl ester ($PC_{61}BM$, 99 % purity) and chlorobenzene (HPLC grade) were purchased from Rieken Metals, Solenne and Sigma-Adrich, respectively. $P3HT:PC_{60}BM$ 2:1 ratio (5:2.5 mg) were dissolved in 0.5 mL chlorobenzene and stirred overnight at 70 °C. The solution was spin-coated at 1 k.rpm directly on top of the HMM substrates. The $P3HT:PC_{60}BM$ blends were then annealed under inert and dry atmosphere at 140 °C for 15 min. UV-vis absorbance and reflectance spectra were measured using a Hitachi U-3310 spectrophotometer and the absorbance was determined with the standard $A = -\log_{10} T$ equation. The reflectivity data were collected with a 10° incident angle.

**Grazing incidence wide angle X-ray scattering**. GIXS measurements were conducted at PLS-II 9A U-SAXS beamline of Pohang Accelerator Laboratory (PAL). From the vacuum undulator (IVU), the X-rays were monochromated with Si(111) double crystals and, with K-B type mirrors, they were focused on the detector. A 2D CCD detector (Rayonix SX165) was used to record the patterns. The sample-to-detector distance was set at 225 mm for energy of 11.08 keV (1.119 Å).

**Transient absorption spectroscopy**. The femtosecond transient absorption measurements carried out in transmittance mode were performed according to previous works,[54] while identical dynamics were confirmed in reflectance mode. Both modes used pump and probe pulses generated by a Ti:Sapphire regenerative amplifier (Spitfire Pro XP, Spectra-Physics) working with an 800 nm output. A Mai Tai laser composed of a mode-locked Titanium-doped sapphire ($Ti^{3+}:Al_2O_3$) laser (Tsunami) and of a diode-pumped continuous wave $Nd:YVO_4$ laser (Millennia). The former was used as the seeding laser for the regenerative amplifier. The latter was used to pump the Tsunami. The regenerative amplifier was based on a Q-switched intra-cavity frequency doubled Nd:YLF laser operating at a repetition rate of 5 kHz and delivered 40 fs long pulse centered at 800 nm. The 325 nm pumping beam was obtained by 4$^{th}$ harmonic generation using two beta barium borate (BBO) crystals after the optical parametric amplifier (TOPAS-Prime, Spectra-Physics) operating at 1300 nm. The pump beam was attenuated to 2.0 mW using neutral density filters in front of the sample. The 725 nm probe beam was selected from a white light continuum probe generated in the visible range with half of the 800 nm regenerative amplifier output power and a 2 mm thickness sapphire crystal. The time delay between the pump and probe beam was varied up to 900 ps using a stable delay line. The time interval for the on-set measurement was ~100 fs in stepping motor and the time interval for the decay measurement was ~10 ps in stepping motor. Pump light was modulated using a mechanical chopper at 220 Hz and the differential transmission $\Delta T/T$ of the probe beam was determined as a function of the delay time with a photodiode and a lock-in detection. A new filter (FSR-RG645, Newport) cutting the light below 640 nm was used to reduce the potential impact of scattered light from the pump beam. Unless stated otherwise, the pump ($\lambda$ = 325 nm, $P$ ~ 2 mW, $\phi$ ~ 1.5 mm, fluence is given in the SI) and probe ($\lambda$ = 725 nm, $P$ ~ 0.01 mW, $\phi$ ~ 100 µm) beams were kept constant and hit the substrates on the dyad-HMM side. The differential transmission was then calculated as

$$\Delta T/T(\lambda,t) = [T_{on}(\lambda,t) - T_{off}(\lambda,t)]/T_{off}(\lambda,t) \quad (8)$$

where $T_{on}$ and $T_{off}$ correspond to the sample transmission with the pump beam on and off, respectively. The charge separation (i.e. onset, rise) and recombination (i.e. decay, recovery) times were measured with 100 fs and 10 ps time interval, respectively. The values were extracted from the $\Delta T/T$ curves plotted as a function of the delay time and fitted with eqs. (9) and (10) for CS and CR, respectively.

$$\Delta T/T(t) = \alpha.\exp(-t/\tau_{CS}) - 1 \quad (9)$$

$$\Delta T/T(t) = -\alpha.\exp(-t/\tau_{CR}) \quad (10)$$

All the dyad transient spectra show a single exponential behaviour, while the P3HT:PCBM data were fitted with a bi-exponential function. The characteristic times obtained in transmittance and reflectance modes are identical within 5 % corresponding to the experimental uncertainty. Considering the setup characteristics, the experimental uncertainty of CS and CR were determined to be 0.05 and 5 ps respectively ; the CT characteristic time errors associated with the fits were always smaller than these values.

**Data availability.** The datasets generated during and/or analysed during the current study are available from the corresponding authors on reasonable request.

**Acknowledgments**

This work has been carried out in the framework of CNRS International Associated Laboratory "Functional nanostructures: morphology, nanoelectronics and ultrafast optics" (LIA NANOFUNC). KJL, YX, JHW, EK, JCR, JWW and PA were supported by funding of the Ministry of Science, ICT & Future Planning, Korea (201000453, 2015001948, 2014M3A6B3063706). PA would like to thank the Canon Foundation in Europe for supporting his Fellowship. The authors also acknowledge the International Research Network (GDRI, CNRS) on "Functional Materials for Organic Optics, Electronics and Devices" (FUNMOOD). The authors are also grateful to Dr J.-Y. Bigot, Dr M. Vomir, Dr. M. Barthelemy and Dr O. Cregut for their contributions in setting-up and optimizing the femtosecond pump-probe setup used in this study. The authors thank Pohang Accelerator Laboratory (PAL) in South Korea for giving us the opportunity to perform the GIXS measurements in the frame of the proposal number "2014-1st-9A-015". The authors are grateful to MEST and POSTECH for supporting the GIXS experiments, to Dr Tae Joo Shin and Dr Hyungju Ahn as well as other staff members from 9A U-SAXS beamline for assistance for adjustments and help. Dr B. Heinrich (BH) is warmly acknowledged for his contributions to the GIXS measurements and data analysis.

**Author contributions**
KJL, JCR, JWW and PA designed the experiments. YX synthesized the dyad molecules under FM, DK and AJA supervision. YX, BH and FM completed the GIXS measurements, which BH analysed. KJL completed the transient measurements and analysis with some assistance of JHW and EK, and with the feedback of PA. KJL, JWW and PA developed the model used to describe the system and discussed the results. KJL completed all the numerical calculations. KJL and PA prepared the figures. PA wrote the manuscript and SI, except the GIXS and dyad preparation. KJL, JCR, JWW and PA replied to the reviewer comments and altered the manuscript accordingly. All the authors could comment the manuscript.

**Additional Information**
Supplementary information is available in the on-line version of the paper: synthesis and product characterisations, grazing incidence wide angle X-ray scattering (GIXS) of the dyad molecules on $Al_2O_3$ interface of an HMM substrate, further steady state and transient optical characterisations, charge transfer model description and further calculations, limits of the solvent analogy. Correspondence and requests should be addressed to PA, JWW and JCR.

**Competing Financial Interests**
The authors declare no competing financial interests.